\documentclass[prx,twocolumn,preprintnumbers,superscriptaddress,longbibliography]{revtex4-1}

\usepackage{amssymb}
\usepackage{amsmath}
\usepackage{amsfonts}
\usepackage{graphicx}
\usepackage{dcolumn}
\usepackage{bm}
\usepackage{verbatim}
\usepackage{color}
\usepackage{wasysym}
\usepackage{xcolor}
\usepackage{times}
\usepackage{soul}
\usepackage{braket}
\usepackage{natbib}
\usepackage{floatrow}
\usepackage{comment}
\usepackage{soul,xcolor}
\usepackage[normalem]{ulem}
\usepackage{ulem}
\usepackage{appendix}
\usepackage[colorlinks,
citecolor = blue,
linkcolor = blue,
urlcolor  = blue,
anchorcolor = blue,
breaklinks = true
]{hyperref}

\definecolor{black}{rgb}{0,0,0}
\definecolor{blue}{rgb}{0,0,1}
\definecolor{green}{rgb}{0,1,0}
\definecolor{red}{rgb}{1,0,0}
\definecolor{brown}{rgb}{0.4,0.2,0}
\definecolor{darkgreen}{rgb}{0,0.7,0}
\definecolor{darkblue}{rgb}{0.0,0.0,0.5}
\definecolor{red}{rgb}{1,0,0}
\definecolor{deepmagenta}{rgb}{0.8, 0.0, 0.8}

\begin{document}

\title{Attractive and Repulsive Angulons in Superfluid Environments}
\date{\today }
\author{Wei Zhang}
\affiliation{Institute of Theoretical Physics, Chinese Academy of Sciences, Beijing 100190, China} 
\affiliation{School of Physical Sciences, University of Chinese Academy of
Sciences, Beijing 100049, China}
\author{Zhongda Zeng}
\affiliation{Institute for Theoretical Physics, University of Innsbruck, Innsbruck A-6020, Austria} 
\affiliation{Institute for Quantum Optics and Quantum Information,
	Austrian Academy of Sciences, Innsbruck A-6020, Austria}
\author{Tao Shi}
\email{tshi@itp.ac.cn}
\affiliation{Institute of Theoretical Physics, Chinese Academy of Sciences, Beijing 100190, China} 
\affiliation{School of Physical Sciences, University of Chinese Academy of
Sciences, Beijing 100049, China}

\begin{abstract}
We investigate the in- and out-of-equilibrium phenomena of a rotational impurity--specifically, a linear molecule--coupled to a nonconventional environment, a helium nanodroplet. By employing a Lee-Low-Pines-like transformation combined with a multireference configuration approach, we self-consistently account for the molecule's backaction on the superfluid bath and accurately capture the complex entanglement between the molecule's rotational degrees of freedom and the bath excitations. Our findings reveal that, in the ground state, the impurity induces a density defect in the superfluid bath, giving rise to two novel types of excited states: (a) attractive angulon states, analogous to bound states in photonic crystals and Yu-Shiba-Rusinov bound states in superconductors, localized within the density defect region; and (b) long-lived repulsive angulon states in dilute environments. Rotational spectroscopy demonstrates a crossover from repulsive to attractive angulon states as the bath density increases. This work paves the way for exploring novel nonequilibrium phenomena of quantum impurities in interacting environments.
\end{abstract}

\maketitle

\section{Introduction}

Quantum impurities coupled to nonconventional baths exhibit novel bound states and non-Markovian dynamics~\cite{Tudela2017,Tudela2018,nonMarkovian2016,nonMarkovian2019}. This has spurred extensive research across
diverse fields, including Kondo physics~\cite%
{10.1143/PTP.32.37,P.B.Wiegmann_1981,PhysRevLett.45.379,ashida2018solving,ashida2018variational}, spin-boson models~\cite{spinboson,ultrastong,Shiultra2018}, central spin models~\cite%
{10.1143/PTP.40.435,doi:10.1143/JPSJ.61.3239,doi:10.1143/JPSJ.62.3181}, polaron problems~\cite%
{landau_electron_1933,Pekar,frohlich_electrons_1954,devreese_frohlich_2015,PhysRev.90.297,PhysRevA.103.063312,dolgirev2021emergence,shi2018variational}, and lattice gauge theories~\cite{PhysRev.82.664,PhysRevD.11.395,LGT2016,sala2018variational}. Of particular interest is the capability of bath excitations to mediate interactions between impurities, leading to intriguing many-body phenomena~\cite{dimmer2014,aainter2015,ShiNJP2018,topo2019}. Molecules embedded in helium nanodroplets represent an exotic impurity-bath system that has garnered significant
attention from both physicists and chemists~\cite%
{toennies_superfluid_2004,yang_helium_2012,Szalewicz01042008,annurev:/content/journals/10.1146/annurev.physchem.49.1.1,10.1063/1.1418746,Choi01012006,Stienkemeier_2006,Mudrich03072014}. The superfluidity of helium effectively suppresses collisional and
Doppler broadening of molecular spectral lines, enabling the stabilization of molecular species (e.g., free radicals) that are otherwise unstable in conventional baths such as gas phases~\cite{10.1063/1.1484104,10.1039/9781782626800-00444}. Consequently, helium nanodroplets serve as an ideal platform for cooling and
manipulating molecules. Studying molecules in such superfluid environments offers a unique opportunity to explore the physics of
impurities interacting with nonconventional baths. Notably,
superfluid environments can host intriguing many-body bound states around the impurity, such as the polaron state in Bose-Einstein condensates (BECs)~\cite%
{PhysRevLett.117.113002} and the Yu-Shiba-Rusinov (YSR) state in
superconductors~\cite{Luh1965BOUNDSI,rusinov1969theory}.

Unlike traditional impurity-bath interacting systems, the rotational motion of molecules interacts with helium atoms, leading to novel dynamical phenomena. For instance, non-trivial shifts and broadenings in the molecular spectrum are observed through rotational spectroscopy~\cite%
{morrison_rotational_2013,10.1063/1.4983703,SLIPCHENKO2005176,toennies_superfluid_2004}. While theoretical models have been proposed to explain these phenomena~%
\cite{schmidt_rotation_2015,schmidt_deformation_2016}, they often predict
anomalous molecular moments of inertia at certain superfluid densities~\cite%
{10.1063/5.0135893}. Numerical methods, including Monte Carlo simulations~\cite%
{zillich_roton-rotation_2004,zillich_quantum_2004,zillich_rotational_2010},
have also been employed to study these spectroscopic phenomena;
however, it remains challenging to circumvent finite-size effects. Therefore, a
unified theory is needed to provide a comprehensive and precise description
of molecular dynamics in superfluid environments across various conditions,
such as varying densities, scattering lengths of host particles, and coupling strengths to molecules.

\begin{figure}[tbp]
\centering				\includegraphics[width=1\textwidth]{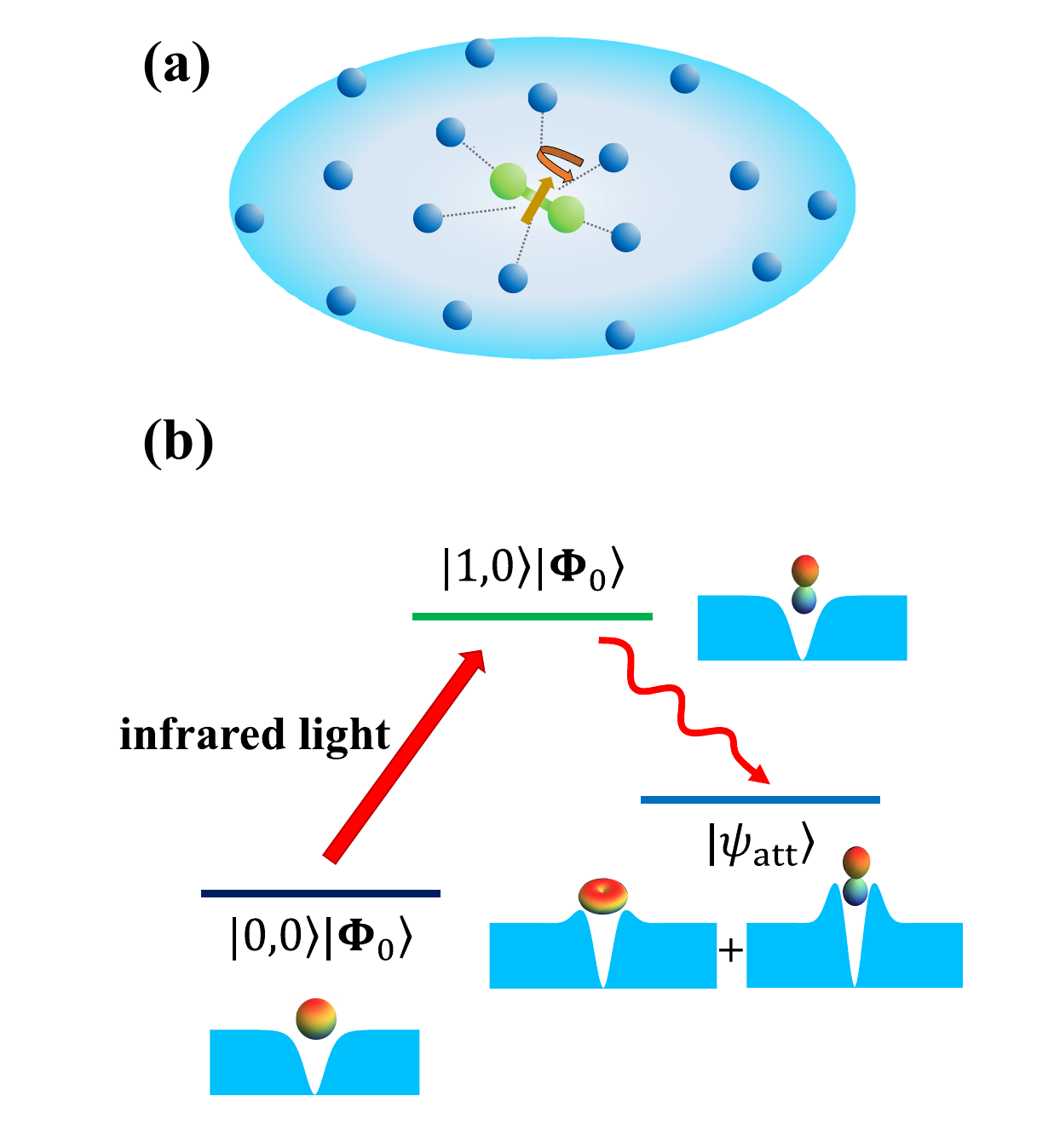}
\caption{Schematics of the impurity-bath system and the rotational spectroscopy process: (a) A linear molecule embedded in a superfluid environment. (b) In its ground state, the molecule induces a superfluid density defect. During rotational spectroscopy, external infrared light excites the molecule to a rotationally excited state. The system then relaxes to the attractive angulon state by emitting Bogoliubov excitations. The attractive angulon state is an entangled configuration of the molecule's rotational state and a density wave localized within the superfluid density defect.}
\label{fig:sketch}
\end{figure}

In this paper, we investigate the in- and out-of-equilibrium physics of a linear molecule in superfluid environments, as schematically shown in Fig.~\ref{fig:sketch}(a), by combining a Lee-Low-Pines (LLP)-like transformation with a multireference configuration approach. Our ansatz accurately captures the entanglement between the molecule's
rotational states and the environmental excitations while
self-consistently accounting for the molecule's backaction on the
superfluid. Through an analysis of the ground state and rotational
spectroscopy, we identify a superfluid density defect created by the
molecule. This defect plays a key role in supporting two novel types of excited states. (a) \textit{Attractive angulon states}: These states form within the density defect at intermediate and high densities and represent the lowest-energy states with fixed nonzero angular momentum. They are analogous to bound states in nonconventional baths~\cite{John1990,Shi2009,Shi2016} and YSR states. (b) \textit{Repulsive angulon states}: These are quasibound states immersed in the continuum of bath excitations. During rotational spectroscopy, the molecule is initially prepared in a rotational excited state by external infrared light, which then spontaneously emits energy into the superfluid bath, as depicted in Fig.~\ref{fig:sketch}(b). By studying the evolution of the molecular impurity and the superfluid density, we observe that at low densities, a long-lived repulsive angulon state persists. In contrast, at intermediate and high densities, the system rapidly relaxes into the attractive angulon state via the emission of Bogoliubov excitations. Furthermore, we demonstrate a density-driven crossover from the repulsive to the attractive angulon state in rotational spectroscopy. Our ansatz also predicts a renormalized moment of inertia for the molecule, which exceeds that of the bare rotor due to the dressing effect from bath excitations, thereby resolving the anomalous moment of inertia renormalization~\cite{10.1063/5.0135893}.

\section{Model}

We consider a linear quantum rotor immersed in a nonconventional bath, specifically interacting Bose gases. The system Hamiltonian $\hat{H}=\hat{H}_{\mathrm{r}}+\hat{H}_{\mathrm{b%
}}+\hat{H}_{\mathrm{r-b}}$ contains three parts~\cite%
{schmidt_rotation_2015,10.1039/9781782626800-00444}. The quantum rotor is
governed by $\hat{H}_{\mathrm{r}}=B{\hat{\bm{J}}}^{2}$, where $B$ is the
rotational constant (the inverse of the moment of inertia). The angular
momentum operators ${\hat{\bm{J}}}=(\hat{J}_{{x}},\hat{J}_{{y}},\hat{J}_{{z}%
})^{\mathrm{T}}$ in the laboratory frame satisfy the $SU(2)$ commutation relations, $[\hat{J}_{{\alpha }},\hat{J}_{{\beta }}]=i\epsilon _{\alpha \beta
\gamma }\hat{J}_{\gamma }$, where $\epsilon _{\alpha \beta \gamma }$ is the
Levi-Civita symbol. The bath Hamiltonian is 
\begin{equation}
\hat{H}_{\mathrm{b}}=\int d^{3}r[\hat{\phi}^{\dagger }(\mathbf{r})(-\frac{%
\nabla ^{2}}{2m_{\mathrm{b}}}-\mu )\hat{\phi}(\mathbf{r})+\frac{g_{\mathrm{bb%
}}}{2}\hat{\phi}^{\dagger 2}(\mathbf{r})\hat{\phi}^{2}(\mathbf{r})],
\end{equation}%
where $\hat{\phi}(\mathbf{r})$ is the annihilation operator for bosons of
mass $m_{\mathrm{b}}$ at position $\mathbf{r}$, $\mu $ is the chemical
potential, and $g_{\mathrm{bb}}=4\pi a_{\mathrm{bb}}/m_{\mathrm{%
b}}$ is the strength of the contact interaction, proportional to the scattering length $a_{%
\mathrm{bb}}$. The interaction between the linear rotor and the Bose gas is
described by%
\begin{equation}
\hat{H}_{\mathrm{r-b}}=\int d^{3}rV_{\mathrm{r-b}}(\mathbf{r},\hat{\theta}_{%
\mathrm{r}},\hat{\varphi}_{\mathrm{r}})\hat{\phi}^{\dagger }(\mathbf{r})\hat{%
\phi}(\mathbf{r}),
\end{equation}%
where $(\hat{\theta}_{\mathrm{r%
}},\hat{\varphi}_{\mathrm{r}})$ are the orientation angle operators of the rotor. The interaction potential is expressed in spherical coordinates $(r,\theta ,\varphi )$ of the host bosons as
\begin{equation}
V_{\mathrm{r-b}}(\mathbf{r},\hat{\theta}_{\mathrm{r}},\hat{\varphi}_{\mathrm{%
r}})=\sum_{{lm}}\sqrt{\frac{4\pi }{2l+1}}Y_{{lm}}^{\ast }(\hat{\theta}_{%
\mathrm{r}},\hat{\varphi}_{\mathrm{r}})Y_{{lm}}(\theta ,\varphi )V_{{l}}(r),
\end{equation}%
where $Y_{{lm%
}}(\theta ,\varphi )$ are spherical harmonics and $V_{l}(r)$ represents the
local radial potential in the $l$-th sector.

Since the entire system is rotationally invariant, we employ a
LLP-like transformation%
\begin{equation}
\hat{S}=e^{i\hat{\gamma}_{\mathrm{r}}\hat{\Lambda}_{z}}e^{i\hat{\theta}_{%
\mathrm{r}}\hat{\Lambda}_{y}}e^{i\hat{\varphi}_{\mathrm{r}}\hat{\Lambda}_{z}}
\end{equation}%
to decouple the total angular momentum $\hat{\bm{J}}^{2}$ of the rotor in
the body-fixed frame~\cite{landau2013quantum,varshalovich1988quantum,schmidt_deformation_2016}. Here, $(\hat{\theta}_{\mathrm{%
r}},\hat{\varphi}_{\mathrm{r}},\hat{\gamma}_{\mathrm{r}})$ are the Euler
angles of the rotor, $\hat{\Lambda}_{\alpha}=\int d^{3}r\hat{\phi}%
^{\dagger }(\mathbf{r})L_{\alpha }\hat{\phi}(\mathbf{r})$ for $\alpha =x,y,z$, and $L_{\alpha }=(\mathbf{r}\times 
\mathbf{p})_{\alpha }$ are the components of the angular momentum operator along the $\alpha $-direction. In the body-fixed
frame, the Hamiltonian becomes%
\begin{equation}
\hat{S}^{\dagger }\hat{H}\hat{S}\equiv \hat{\mathcal{H}}=\hat{\mathcal{H}}_{%
\mathrm{r}}+\hat{{H}}_{\mathrm{b}}+\hat{\mathcal{H}}_{\mathrm{r-b}},
\end{equation}%
where $\hat{\mathcal{H}}_{\mathrm{r}}=B(\hat{\bm{\mathcal{J}}}-\hat{\mathbf{%
\Lambda }})^{2}$, and the generators $\hat{\bm{\mathcal{J}}}$ of body
rotations satisfy $[\hat{\mathcal{J}}_{\alpha },%
\hat{\mathcal{J}}_{\beta }]=-i\epsilon _{\alpha \beta \gamma }\hat{\mathcal{J%
}}_{\gamma }$~\cite{schmidt_deformation_2016, PhysRevD.11.395}. The rotor-bath interaction term becomes
\begin{equation}
\hat{\mathcal{H}}_{\mathrm{r-b}}=\int d^{3}r\sum_{l}V_{l}(r)Y_{l0}(\theta
,\varphi )\hat{\phi}^{\dagger }(\mathbf{r})\hat{\phi}(\mathbf{r}),
\end{equation}%
and it is independent of the rotor operators $(\hat{\theta}_{\mathrm{r}},%
\hat{\varphi}_{\mathrm{r}})$.

It is important to note that ${\hat{\bm{\mathcal{J}}^{2}}}=\hat{\bm{J}}^{2}$ and $\hat{%
\mathcal{M}}_{z}=\hat{\mathcal{J}}_{z}-\hat{\Lambda}_{z}$ are conserved quantities under $%
\hat{\mathcal{H}}$. Therefore, the dynamics can be studied independently in
different sectors with fixed $(\mathcal{J},\mathcal{M}_{z})$. However, due
to the non-trivial cross term $\hat{\bm{\mathcal{J}}}\cdot \hat{\mathbf{%
\Lambda }}$ in $\hat{\mathcal{H}_{\mathrm{r}}}$, each component $\hat{%
\mathcal{J}}_{\alpha }$ does not commute with $\hat{\mathcal{H}}$ in the $%
\mathcal{J}>0$ sector. Consequently, the transformation $\hat{S}$ cannot
fully decouple the rotor degrees of freedom, and the entanglement between
the rotor and bath particles for $\mathcal{J}>0$ must be carefully treated.

\section{Formalism}

We study the ground state properties and real-time dynamics governed by the
Hamiltonian $\hat{\mathcal{H}}$ using a multireference configuration
approach~\cite{ROOS1980157,10.1063/1.455063,doi:10.1021/j100377a011}. In the 
$\mathcal{J}$-th sector, we consider a variational state in the LLP frame:
\begin{equation}
|\psi ^{\mathcal{J}}\rangle =\sum_{M=-\mathcal{J}}^{\mathcal{J}}c_{M}|%
\mathcal{J},{M}\rangle |f_{M}\rangle.  \label{var_ansatz}
\end{equation}%
This state is a superposition of product states $|\mathcal{J},{M}\rangle |f_{M}\rangle $, where $c_{M}$ are the superposition coefficients, $|\mathcal{J},M\rangle $
denote common eigenstates of the rotor operators $\hat{\bm{\mathcal{J}}^{2}}$ and $\hat{\mathcal{%
J}_{z}}$, and the coherent state%
\begin{equation}
|f_{M}\rangle =\exp [\int d^{3}rf_{M}(\mathbf{r})\hat{\phi}^{\dagger }(%
\mathbf{r})-\mathrm{H.c.}]|0\rangle  \label{var_ansatz1}
\end{equation}%
describes the condensate in the bath with the spatial wavefunction $f_{M}(\mathbf{r})$.
In the limit $B\rightarrow 0$, $\hat{\mathcal{H}_{\mathrm{r}}}$ vanishes, and
all $\mathcal{J}_{\alpha }$ are conserved. Thus, the state $|\psi ^{\mathcal{J%
}}\rangle $ reduces to a product state consisting of $|\mathcal{J},M\rangle $
and a single coherent state for bath bosons. For finite $B$, the state $%
|\psi ^{\mathcal{J}}\rangle $ effectively captures the entanglement between
the rotor and bath bosons.

To obtain the equations of motion (EOM) for the variational parameters $c_{M}$ and $f_{M}(\mathbf{r})$, we project the imaginary time evolution equation
\begin{equation}
\partial _{\tau }|\psi ^{\mathcal{J}}(\tau )\rangle =-(
\hat{\mathcal{H}}-E)|\psi ^{\mathcal{J}}(\tau )\rangle\label{IM}
\end{equation}
and the Schr\"{o}dinger
equation
\begin{equation}
i\partial _{t}|\psi ^{\mathcal{J}}(t)\rangle=\hat{%
\mathcal{H}}|\psi ^{\mathcal{J}}(t)\rangle 
\label{SE}
\end{equation}
onto the tangent space spanned by the vectors 
$\partial |\psi^{\mathcal{J}}\rangle/ \partial c_M$ and $\delta |\psi^{\mathcal{J}}\rangle/ \delta f_M$ of the variational manifold $|\psi ^{\mathcal{J}}\rangle$~\cite{shi2018variational,shi2020variational} . Here, the variational energy $E=\langle
\psi ^{\mathcal{J}}|\hat{\mathcal{H}}|\psi ^{\mathcal{J}}\rangle $ monotonically decreases during imaginary time evolution and remains conserved during real-time evolution. The analytical expressions of EOM for the variational parameters are shown in Appendix A. We further expand the condensate wavefunctions as ${f}_{M}(\mathbf{r})=\sum_{l,m}Y_{lm}(\theta ,\phi ){f}_{M,lm}(r)$ in the angular momentum basis and numerically solve the resulting EOM using the Hankel transformation~\cite{PhysRevA.74.013623,PhysRevResearch.2.043074,PhysRevResearch.4.043018,Pan_2022}. In our numerical calculations, we set the scattering length $a_{\mathrm{bb}%
}=3.3(mB)^{-1/2}$, corresponding to the speed of sound in superfluid He$%
^{4}$ with $B=2\pi \times 1$GHz. We also use the effective potential~\cite%
{schmidt_rotation_2015} $%
V_{l}(r)=(2\pi )^{-3/2}u_{l}e^{-r^{2}/2r_{l}^{2}}$ with strengths $u_{l}$
and ranges $r_{l}$ for channels $l=0,1,...,l_{c}$, where $l_{c}$ is the
angular momentum cutoff. For instance, we take $r_{0}=r_{1}=1.5(m_{\mathrm{b}%
}B)^{-1/2}$ and $u_{0}=1.75u_{1}=218B$ for $l=0$ and $1$, corresponding to a
typical atom-molecule interaction potential~\cite%
{schmidt_rotation_2015,stone2013theory,10.1063/1.556028}. Throughout the paper, we use $B$, $(m_{b}B)^{-1/2}$, and $%
(m_{b}B)^{3/2}$ as the units of energy, length, and density, respectively.

\section{Ground states and attractive angulons}

\begin{figure}[tbp]
\centering		
\includegraphics[width=1.0\textwidth]{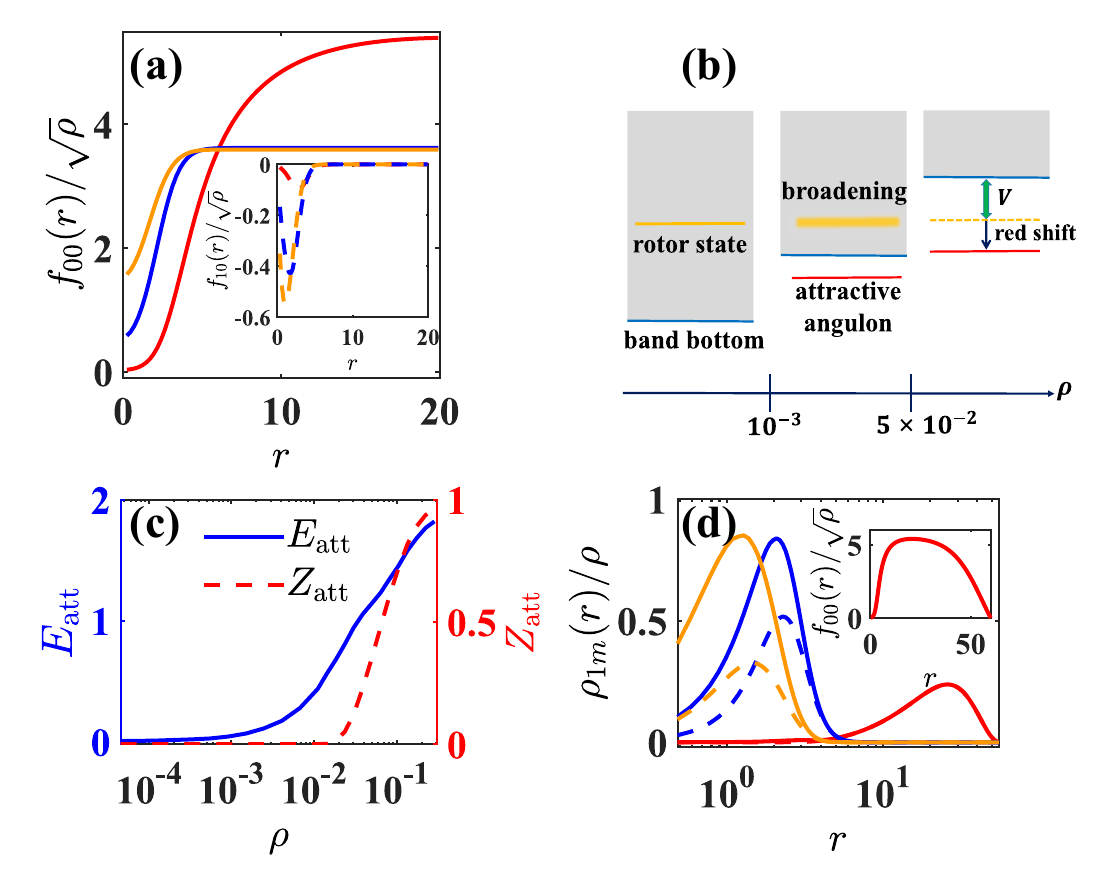}
\caption{
(a) Condensate wavefunction in the ground state,  $f_{00}(r)/\protect\sqrt{\rho}$
(solid curve) and $f_{10}(r)/\protect\sqrt{\rho}$ (dashed curve) at different
densities: $\protect\rho=5\times10^{-5}$ (red), $3.5\times10^{-2}$
(blue), $8\times10^{-2}$ (orange). 
(b) The schematic of energy
spectra for finite $\mathcal{J}$ in different density regimes, where the
gray regions and the orange lines denote the continuum of Bogoliubov excitations and the bare
rotor state, respectively. As the density increases, the bottom of the Bogoliubov excitation
continuum shifts upward, rising from $3.2 \times 10^{-3}$ at $\protect\rho=5\times10^{-5}$, to $1.028$ at $\protect\rho=3.5\times10^{-2}$, and further to $2.004$ at $\protect\rho=8\times10^{-2}$.
The hybridization strength between the rotor
state and continuum is enhanced by the density, resulting in the emergence of the
attractive angulon state and a broadening of the rotor state. 
(c) The energy
(solid blue curve) and the single particle residue (dashed red curve) of the
attractive angulon state in the $\mathcal{J}=1$ sector as functions of the
density. 
(d) Density distributions $\protect\rho_{10}(r)/\protect\rho$
(solid curve) and $\protect\rho_{11}(r)/\protect\rho=\protect\rho_{1-1}(r)/%
\protect\rho$ (dashed curve) in the attractive angulon state at different
densities: $\protect\rho=5\times10^{-5}$ (red), $3.5\times10^{-2}$
(blue), $8\times10^{-2}$ (orange). When the density is low, $%
\protect\rho_{10}(r)$ almost coincides with $\protect\rho_{11}(r)=\protect%
\rho_{1-1}(r)$.}
\label{fig:FIG1}
\end{figure}

The global ground state resides in the $\mathcal{J}=0$ sector in the
body-fixed frame. Since the transformation $\hat{S}$ maps the total angular
momentum to $\mathcal{J}$, the ground state is rotationally invariant in the
laboratory frame. In the asymptotic limit $\tau \rightarrow \infty$, the
fixed-point solution of Eq.~\eqref{IM} yields the ground state $|\psi _{%
\mathrm{GS}}\rangle =|0,0\rangle |\Phi _{0}\rangle $, with ground state
energy $E_{\mathrm{GS}}=\langle \psi _{\mathrm{GS}}|\hat{\mathcal{H}}|\psi _{%
\mathrm{GS}}\rangle $. Here, the coherent state%
\begin{equation}
|\Phi _{0}\rangle =\exp [\int d^{3}rf_{\mathrm{GS}}(\mathbf{r})\hat{\phi}%
^{\dagger }(\mathbf{r})-\mathrm{H.c.}]|0\rangle
\end{equation}%
describes a non-uniform condensate with wavefunction $f_{\mathrm{GS}}(%
\mathbf{r})$. In Fig.~\ref{fig:FIG1}(a) and its inset, the components ${f}_{%
\mathrm{GS},lm}(r)=\int d\Omega _{\mathbf{r}}Y_{lm}^{\ast }(\Omega _{\mathbf{%
r}})f_{\mathrm{GS}}(\mathbf{r})$ display predominant condensation in the $s$%
-wave channel $(l,m)=(0,0)$, with a significantly smaller occupation in the $p$-wave
channel $(l,m)=(1,0)$, where $d\Omega _{\mathbf{r}}=\sin \theta d\theta
d\varphi $. The rotor induces a repulsive potential $V_{l}(r)$ for the
bath bosons, leading to the formation of a density defect with a healing length $\xi =1/\sqrt{8\pi \rho a_{\mathrm{%
bb}}}$ in the condensate. This defect plays a crucial role in
generating intriguing bound states as the density increases.

With respect to the Gaussian ground state $|\psi _{\mathrm{GS}}\rangle $, a
quadratic mean-field Hamiltonian $\hat{\mathcal{H}}_{\mathrm{MF}}$ can be
constructed using Wick's theorem in the $\mathcal{J}=0$ sector, as detailed in
Appendix B. The eigenstates and eigenvalues of $\hat{\mathcal{H}}_{\mathrm{MF%
}}$ correspond to the Bogoliubov excitations, denoted as $\hat{b}%
_{\alpha m}^{\dag }|\Phi _{0}\rangle $, and their associated spectrum. The annihilation
operator $\hat{b}_{\alpha m}$ of a Bogoliubov excitation is a superposition of $\delta \hat{a}%
_{lm}(r)=\hat{a}_{lm}(r)-{f}_{\mathrm{GS},lm}(r)$ and $\delta \hat{a}%
_{lm}^{\dagger }(r)$, where $\hat{a}_{lm}(r)=\int
d\Omega _{\mathbf{r}}Y_{lm}^{\ast }(\Omega _{\mathbf{r}})\hat{\phi}(\mathbf{r%
})$ is the annihilation operator in the channel $(l,m)$. Due to the
cylindrical symmetry of the linear rotor, the operator $\hat{b}_{\alpha m}$
has a well-defined projected angular momentum $m$ along the $z$-direction, where $\alpha $
labels the eigenmode with fluctuations along the radial and polar-angle
directions. The spectrum of Bogoliubov excitations is depicted by the gray regions in Fig.~\ref%
{fig:FIG1}(b), where the bottom of the spectrum shifts upward as the density
increases.

To explore the ground state in the $\mathcal{J}>0$ sector, we construct an
effective Hamiltonian $H_{\mathrm{eff}}=\hat{P}\hat{\mathcal{H}}\hat{P}$
using the projection operator $\hat{P}$ onto the subspace $\mathcal{S}%
=\{|\Xi ^{\mathcal{J}}\rangle ,\left\vert \mathcal{J},m\right\rangle \hat{b}%
_{\alpha m}^{\dag }|\Phi _{0}\rangle \}$. The first state $|\Xi ^{\mathcal{J}%
}\rangle =|\mathcal{J},0\rangle |\Phi _{0}\rangle $ in $\mathcal{S}$
describes a bare rotor excitation on top of the ground-state condensate,
where $\hat{\mathcal{M}}_{z}|\Xi ^{\mathcal{J}}\rangle =0$. Due to the
conservation of $\hat{\mathcal{M}}_{z}$, the Hamiltonian $\hat{\mathcal{H}}$
hybridizes $|\Xi ^{\mathcal{J}}\rangle $ and the continuum states $\left\vert 
\mathcal{J},m\right\rangle \hat{b}_{\alpha m}^{\dag }|\Phi _{0}\rangle $ while conserving $\mathcal{M}_{z} =0$. As
illustrated in Fig.~\ref{fig:FIG1}(b), $H_{\mathrm{eff}}$ describes how a single
energy level $|\Xi ^{\mathcal{J}}\rangle $ couples resonantly and off-resonantly to the Bogoliubov excitation continuum across different density regimes, giving rise to novel
non-Markovian dynamics~\cite{John1990,ultra2014,gonzaleztudela17b}.

Diagonalizing $H_{\mathrm{eff}}$ yields the eigenstates%
\begin{equation}
|\psi _{n}^{\mathcal{J}}\rangle =\sqrt{Z_{n}}|\Xi ^{\mathcal{J}}\rangle
+\sum_{\alpha m}\psi _{n,\alpha m}\left\vert \mathcal{J},m\right\rangle \hat{%
b}_{\alpha m}^{\dag }|\Phi _{0}\rangle  \label{CA}
\end{equation}%
and corresponding eigenenergies $E_{n}^{\mathcal{J}}$ ordered in
ascending energy. Notably, although Eq.~\eqref{CA}
resembles the Chevy ansatz (CA)~\cite%
{schmidt_rotation_2015,schmidt_deformation_2016,10.1039/9781782626800-00444,10.1063/5.0135893,PhysRevA.74.063628}, there are two key advantages: (i) we focus on the ground state $|\psi
_{n=0}^{\mathcal{J}}\rangle $, which is not captured by the conventional CA
in low- and intermediate-density regimes~\cite{10.1063/5.0135893}; and
(ii) more significantly, the nonuniform condensate described by $|\Phi
_{0}\rangle $ exhibits a density defect created by the rotor, forming a
trapping potential for Bogoliubov excitations. We will show that the ground state $|\psi
_{n=0}^{\mathcal{J}}\rangle $ is analogous to the attractive polaron state
for a mobile impurity in BECs~\cite{PhysRevLett.117.113002,PhysRevLett.117.055302}. Hence we refer
to it as the \textit{attractive angulon state} $|\psi _{\mathrm{att}}\rangle \equiv
|\psi _{n=0}^{\mathcal{J}}\rangle $. In Fig.~\ref{fig:FIG1}(c), we display the
attractive angulon energy $E_{\mathrm{att}}\equiv E_{n=0}^{\mathcal{J}=1}$
and the single particle residue $Z_{\mathrm{att}}\equiv Z_{n=0}^{\mathcal{J}%
=1}$ as solid (blue) and dashed (red) curves, respectively.

In the dilute regime $\rho \lesssim 1\times 10^{-3}$, the energy $E_{\mathrm{%
att}}$ (solid blue curve) increases monotonically from zero as $\rho $
rises. The tiny residue $Z_{\mathrm{att}}\lesssim 10^{-9}$ (dashed red
curve) indicates that the state $|\psi _{\mathrm{att}}\rangle $ is a
superposition of $\hat{b}_{\alpha m}^{\dag }|\Phi _{0}\rangle $, exhibiting
strong entanglement of rotor states $|1m\rangle $ and Bogoliubov excitations near the bottom of
the continuum in the body-fixed frame. The red curve in Fig.~\ref{fig:FIG1}%
(d) shows the density distribution $\rho _{lm}(r)\equiv \langle \psi _{%
\mathrm{att}}|\hat{a}_{lm}^{\dag }(r)\hat{a}_{lm}(r)|\psi _{\mathrm{att}%
}\rangle $ in the angular momentum channel ($l$,$m$), where $
\protect\rho_{10}(r)$ almost coincides with $\protect\rho_{11}(r)=\protect%
\rho_{1-1}(r)$. At low densities, the
extended density distribution $\rho _{1m}(r)$ over the system suggests that
in the attractive angulon state the Bogoliubov excitation rotates around the rotor with orbital
angular momentum $l=1$ and low kinetic energy. In our numerical
calculations, a hard-wall potential at radius $R=60$ is adopted, causing the
background condensate wavefunction $f_{\mathrm{GS},00}(r)$ to drop rapidly
at the boundary, as shown in the inset of Fig.~\ref{fig:FIG1}(d). This
defect results in the density distribution $\rho _{1m}(r)$ being localized
at the edge, which is merely a boundary effect. Furthermore, $\langle (\hat{%
\bm{\mathcal{J}}}-\hat{\mathbf{\Lambda }})^{2}\rangle =0$ indicates perfect
screening of the rotor's angular momentum by the excitation cloud in the body-fixed
frame. However, in the laboratory frame, the attractive angulon state simplifies to
a trivial product state $\hat{S}|\psi _{\mathrm{att}}\rangle =|00\rangle
|\Phi _{\mathrm{att}}\rangle $, where
\begin{eqnarray}
|\Phi _{\mathrm{att}}\rangle&=&\frac{1}{\sqrt{3}}\sum_{m,m^{\prime}=-1}^1 \int d\gamma_{\rm r} d\theta_{\rm r}\sin\theta_{\rm r} d\varphi_{\rm r}\\ \notag
&\ &Y^*_{00}(\theta_{\rm r},\varphi_{\rm r})[Y_{1m^{\prime}}(\theta_{\rm r},\varphi_{\rm r})D^1_{m,m^{\prime}}(\gamma_{\rm r},\theta_{\rm r},\varphi_{\rm r})]\delta\hat{a}_{k_01m}^{\dag}|\Phi_0\rangle.
\label{phiatt}
\end{eqnarray}
is determined by the Wigner $D$-matrix $D^l_{m,m^{\prime}}(\gamma_{\rm r},\theta_{\rm r},\varphi_{\rm r})$, the minimal quantized momentum $k_0=\alpha _{1}^{(1)}/R$ in the nanodroplet of size $R$, and $\alpha _{1}^{(1)}$ is the first zero of the spherical Bessel function $j_{1}(x)$.

In the intermediate density regime $1\times 10^{-3}\lesssim \rho \lesssim
5\times 10^{-2}$, the lower bound of the Bogoliubov excitation continuum shifts towards the
rotor energy $2B$, as shown in Fig.~\ref{fig:FIG1}(b). Simultaneously, the
coupling strength (proportional to $\sqrt{\rho }$) between $|\Xi ^{\mathcal{J}%
=1}\rangle $ and the continuum states $\hat{b}_{\alpha m}^{\dag }|\Phi
_{0}\rangle $ increase with density. Consequently, the attractive angulon
state emerges below the bottom of the continuum. Figure~\ref{fig:FIG1}(c)
illustrates that as the density increases, the energy $E_{\mathrm{att}}$ approaches 
$2B$ with an increasing occupation $Z_{\mathrm{att}}$. In Fig.~\ref{fig:FIG1}%
(d), the density distributions $\rho _{1m}(r)$ reveal that in the attractive
angulon state, Bogoliubov excitations with angular momentum $l=1$ form a bound state localized around the rotor, where the localization length is approximately the healing length. This attractive angulon state is analogous to bound states in photonic crystals~\cite{John1990,Shi2009,Shi2016} and YSR states in
superconductors~\cite{Luh1965BOUNDSI,10.1143/PTP.40.435,rusinov1969theory}.
The emergence of the bound state can be understood as follows: as the
density increases, the defect of size $\xi $ deepens, as shown in Fig.~\ref%
{fig:FIG1}(a), producing a stronger attractive potential for Bogoliubov excitations. Once this
attractive interaction exceeds a certain threshold, the bound state appears
within the defect potential, indicating that the system has crossed a
shape-scattering resonance~\cite{Blume2008}. Unlike the conventional CA, the non-uniform
condensate background in Eq.~\eqref{CA} incorporates the rotor's backaction,
which is crucial for generating the attractive angulon state.

In the high-density regime $\rho >5\times 10^{-2}$, the bottom of the Bogoliubov excitation
continuum surpasses the rotor energy $2B$ (see Fig.~\ref{fig:FIG1}(b)).
Their coupling leads to a redshift of the rotor state, forming the
attractive angulon state with a significant component $Z_{\mathrm{att}}>0.5$ (see Fig.~\ref{fig:FIG1}(c)). The corresponding density profiles $\rho_{1m}(r)$ in Fig.~\ref{fig:FIG1}(d) show the attractive angulon state localized inside the defect potential, further demonstrating the
characteristics of the bound state. At higher densities, the healing length
decreases, resulting in a more localized attractive angulon state.

\section{Rotational spectroscopy}

Excited states, including attractive angulons, can be experimentally
detected using rotational spectroscopy~\cite{10.1063/1.1418746,grebenev_superfluidity_1998,Stienkemeier_2006,PhysRevLett.76.4560}. For the
system prepared in the ground state $|\psi _{\mathrm{GS}}\rangle $, the
rotor is abruptly excited to a high-angular-momentum state $|\mathcal{J},
0\rangle $ via external infrared light. Subsequently, the initial state $%
\left\vert \psi (0)\right\rangle =|\Xi ^{\mathcal{J}}\rangle $ relaxes into
the attractive angulon state by emitting Bogoliubov excitations into the bath. This process is schematically displayed in Fig.~\ref{fig:sketch}(b).

The real-time dynamics is governed by Eq.~\eqref{SE}. The evolution of
variational parameters $c_{M}$ and $f_{M}(\mathbf{r})$ in $|\psi ^{\mathcal{J%
}}(t)\rangle $ is numerically achieved by solving the EOMs detailed in Appendix A. From the resulting $%
|\psi ^{\mathcal{J}}(t)\rangle $, we compute the spectral function $A(\omega
)=-\mathrm{Im}G(\omega )/\pi $ via the Fourier transform of the retarded
Green function $G(t)=-i\theta (t)e^{iE_{\mathrm{GS}}t}\langle \psi (0)|\psi
^{\mathcal{J}}(t)\rangle $, where $\theta (t)$ is the Heaviside function.
The spectral function reveals visible weight at frequencies corresponding to
the excited states in response to the external light. Without loss of
generality, we focus on the spectrum in the $\mathcal{J}=1$ sector below.

\begin{figure}[tbp]
\centering				\includegraphics[width=1\textwidth]{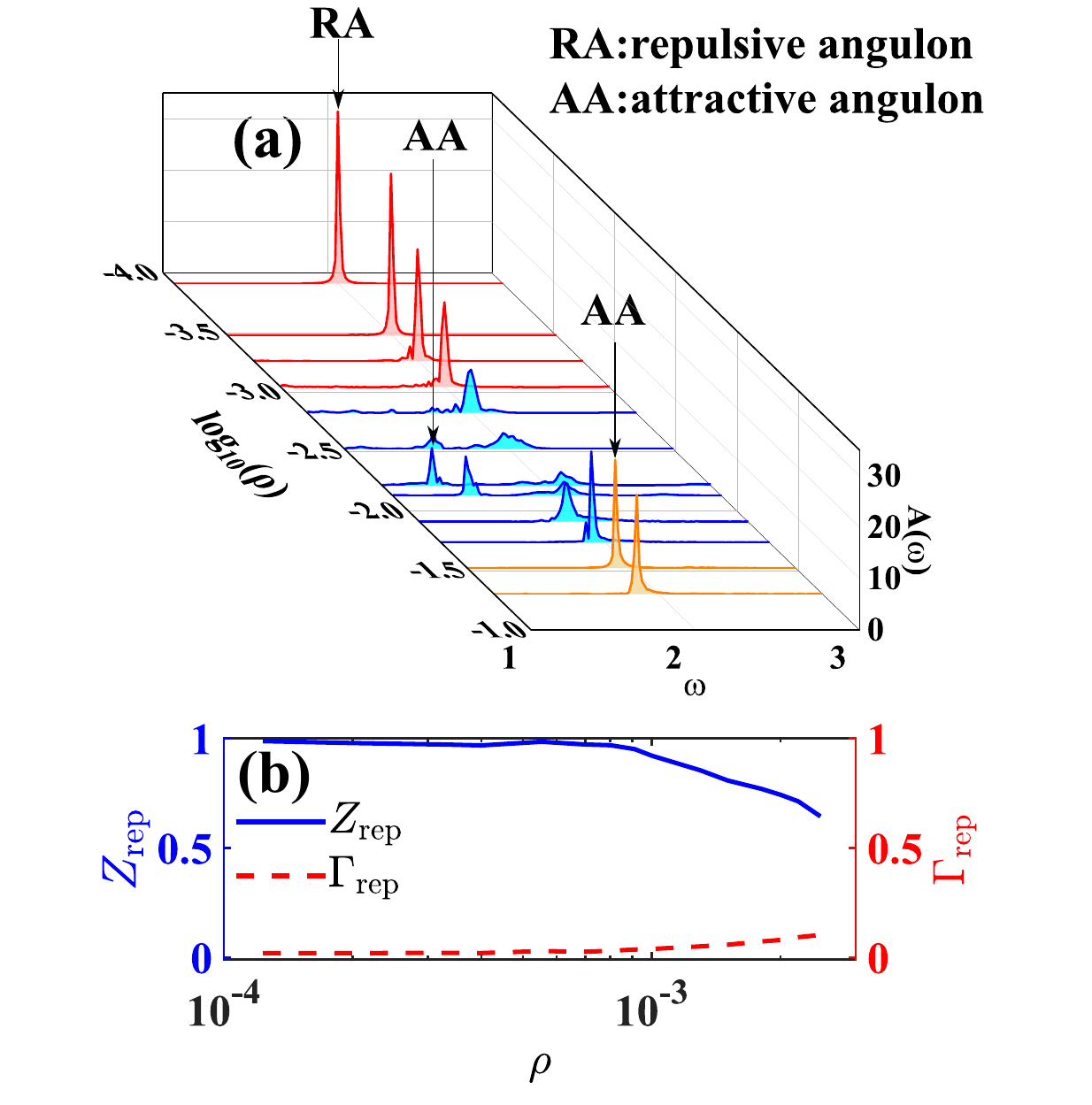}
\caption{(a) Rotational spectroscopy obtained using the multireference
state at varying densities. The spectrum lines in the dilute, intermediate,
and high-density regimes are labeled in red, blue, and orange, respectively. (b) The weight $Z_{\mathrm{rep}}$ and the width $\Gamma_{\mathrm{rep}%
}$ of the repulsive angulon in the low-density regime, obtained by fitting
the rotational spectroscopy to a Lorentzian line shape.}
\label{fig:spectrum_SCS}
\end{figure}

In the dilute regime, the initial rotor state is immersed in the Bogoliubov excitation
continuum (see Fig.~\ref{fig:FIG1}(b)). Here, weak coupling leads to a
slow spontaneous emission, causing the initial state to relax
to a long-lived meta-stable state, referred to as the \textit{repulsive angulon state} 
$|\psi _{\mathrm{rep}}\rangle $. The rotational spectrum exhibits a single
sharp peak corresponding to $|\psi _{\mathrm{rep}}\rangle $, labeled as
\textquotedblleft $\mathrm{RA}$\textquotedblright\ in Fig.~\ref%
{fig:spectrum_SCS}(a). From the position, width, and spectral weight of this
peak, we can extract the energy $E_{\mathrm{rep}}$, decay rate $\Gamma _{%
\mathrm{rep}}$, and the single-particle residue $Z_{\mathrm{rep}}=|\langle
\Xi ^{\mathcal{J}=1}|\psi _{\mathrm{rep}}\rangle |^{2}$ of the repulsive
angulon. We find that the repulsive angulon energy $E_{\mathrm{rep}}\sim
1.99B$ is slightly lower than the rotor energy, which remains largely unaffected in the
dilute density regime. In Fig.~\ref%
{fig:spectrum_SCS}(b), the large single-particle
residue $Z_{\mathrm{rep}}$ and small $\Gamma _{\mathrm{rep}}$ indicate the
long-lived nature of the repulsive angulon due to weak hybridization with Bogoliubov excitations. The single
peak structure is consistent with the previous studies~\cite%
{schmidt_rotation_2015,10.1063/5.0135893}, however, it corresponds to
the repulsive angulon state rather than the true ground state in the sector $%
\mathcal{J}>0$, i.e., the attractive angulon state $|\psi _{\mathrm{att}%
}\rangle $. The absence of the attractive angulon peak in the rotational
spectrum is consistent with its negligible residue $Z_{\mathrm{att}}$ shown
in Fig.~\ref{fig:FIG1}(c) for dilute gases.

\begin{figure}[tbp]
\centering		
\begin{minipage}[t]{1\linewidth}	
			\centering				\includegraphics[width=1\textwidth]{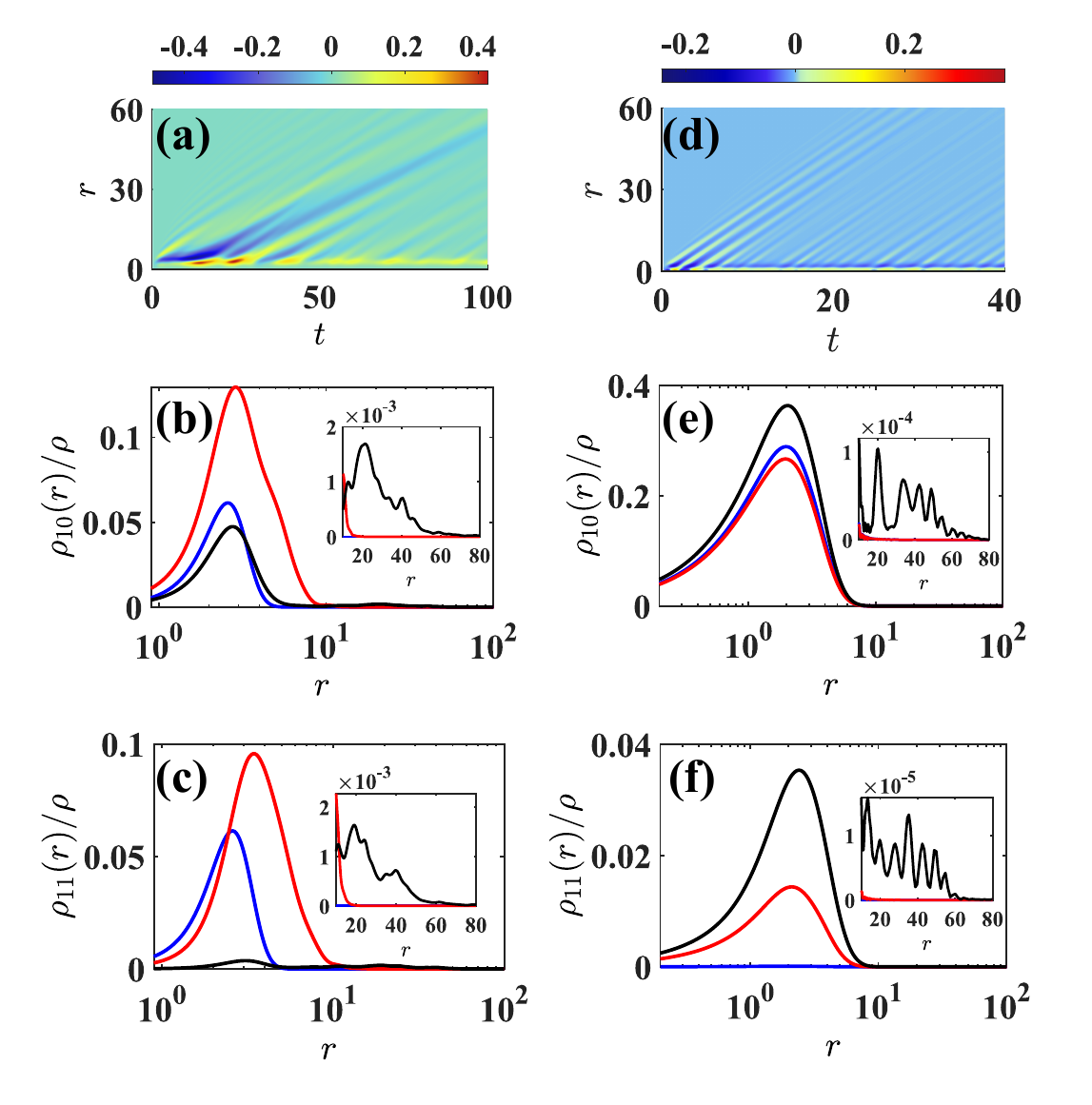}			
		\end{minipage} 
\caption{
Left and right panels correspond to $\rho =6\times 10^{-3}$ and $\rho =9\times 10^{-2}$, respectively. (a),(d) Evolution of density fluctuation $\delta \rho (r,t)=[\rho (r,t)-\rho (r,0)]/\rho$. (b),(c),(e),(f) Boson density distributions for angular momenta $(l,m)=(1,0)$ and $(1,1)$ at $t=1$, $10$, and $100$ (left: blue, red, and black, respectively) and $t=0.01$, $0.2$, and $20$ (right: blue, red, and black, respectively).
}
\label{fig:density_fluc}
\end{figure}

As the density increases to the intermediate regime, the stronger coupling
leads to a faster decay of the initial state $|\psi (0)\rangle $, resulting in
a significantly broadened peak corresponding to the short-lived repulsive
angulon, as shown by the blue curves in Fig.~\ref{fig:spectrum_SCS}(a) (cf.~Fig.~\ref{fig:FIG1}(b)). Additionally, another sharp peak labeled as
\textquotedblleft $\mathrm{AA}$\textquotedblright\ emerges, representing the
attractive angulon state. In this regime, the attractive angulon becomes
visible in the spectrum due to the finite residue $Z_{\mathrm{att}}$. We note that the frequency of the AA peak in the rotational spectrum is higher than the value predicted by the single-excitation ansatz~\eqref{CA}. This discrepancy arises because the multireference configuration ansatz (cf.~Eq.~\eqref{var_ansatz}) accounts for interactions among Bogoliubov excitations. Specifically, the repulsive interactions among multiple Bogoliubov excitations within the angulon cloud increase the attractive angulon energy. During the
relaxation process, the initial state decays to lower energy intermediate
states by successively emitting Bogoliubov excitations, as indicated by satellite peaks in
between $\mathrm{AA}$ and the rotor energy $2B$. Eventually, the system
relaxes to the stable attractive angulon state. We illustrate this process
using the time evolution of the density fluctuation%
\begin{eqnarray}
\delta \rho (r,t) &=&\frac{1}{\rho }[\rho (r,t)-\rho (r,0)],  \notag \\
\rho (r,t) &=&\int d\Omega _{\mathbf{r}}\langle \psi ^{\mathcal{J}}(t)|\hat{%
\phi}^{\dag }(\mathbf{r})\hat{\phi}(\mathbf{r})|\psi ^{\mathcal{J}%
}(t)\rangle ,
\end{eqnarray}
and the density distribution 
\begin{equation}
\rho _{lm}(r,t)\equiv \langle \psi ^{\mathcal{J%
}}(t)|\hat{a}_{lm}^{\dag }(r)\hat{a}_{lm}(r)|\psi ^{\mathcal{J}}(t))\rangle
\end{equation}
in the channel $(l,m)$. As shown in Fig.~\ref{fig:density_fluc}(a), $\delta \rho (r,t)$ reveals
the generation of multiple Bogoliubov excitations, which propagate towards the boundary of the system. In Figs.~\ref{fig:density_fluc}(b) and~\ref{fig:density_fluc}(c), $%
\rho _{lm}(r,t)$ at different instants $t=1$, $10$, and $100$ show that the
density distribution around the rotor gradually stabilizes into that of the
attractive angulon state, where the wavepackets at large $r$ depicted in the insets describe
out-going Bogoliubov excitations. Notably, the strong entanglement between the rotor and the bath excitations during the relaxation process cannot be captured by a single coherent state ansatz~\cite{10.1063/5.0135893}.

In the high-density regime, a red-shifted sharp peak corresponding to the
attractive angulon state appears, while the repulsive angulon completely vanishes in the spectrum, as illustrated by the orange curves in Fig.~\ref%
{fig:spectrum_SCS}(a). For a typical density $\rho =9\times 10^{-2}$, the
peak frequency $\omega \sim 1.9B$ is below the rotor energy, which
quantitatively agrees with the attractive angulon energy depicted in Fig.~%
\ref{fig:FIG1}(c). The high density leads to a significant overlap $Z_{
\mathrm{att}}$ and a short healing time $t_{{\mathrm{healing}}}\sim 1/(\rho
a_{\mathrm{bb}})$, resulting in a rapid relaxation into the attractive angulon
state. The density evolution is shown by $\delta \rho (r,t)$ and $\rho
_{lm}(r,t)$ in Figs.~\ref{fig:density_fluc}(d)--(f). Due to the large overlap 
$Z_{\mathrm{att}}$, only a few Bogoliubov excitations are emitted, allowing the density
distribution around the rotor to quickly stabilize into that of the
attractive angulon state, as shown in Figs.~\ref{fig:density_fluc}(e) and ~%
\ref{fig:density_fluc}(f) for $t=0.01$, $0.2$, and $20$.

\section{Discussion of the effective Rotational Constant}
In previous studies, analogous to the effective mass in polaron problems,
the concept of an \textit{effective rotational constant} (or equivalently, the
rotational Lamb shift) was introduced to describe the renormalization effect
on molecular rotation induced by a bosonic bath~\cite%
{schmidt_rotation_2015,PhysRevLett.118.095301,10.1063/5.0135893}.
This constant, denoted by ${B}_{\mathcal{J}}^{\ast }$, is defined as:%
\begin{equation}
\begin{aligned}
{B}^*_{\mathcal{J}}=\frac{E_{\mathcal{J}}-E_0}{\mathcal{J}(\mathcal{J}+1)},
\mathcal{J}=1,2,..., \end{aligned}
\end{equation}%
where $E_{\mathcal{J}}$ denotes the lowest energy in the $\mathcal{J}$
sector. However, the Chevy ansatz yields ${B}_{1}^{\ast }>B$ in the
intermediate density regime~\cite{10.1063/5.0135893}, which is
nonphysical since the moment of inertia of the rotor should be enhanced by
the dressing effect. In contrast, our analysis reveals that the conventional
Chevy ansatz may be invalid in the intermediate density regime since it does not capture the
density defect in the BEC induced by the molecule's backaction. Our result shows that in the dilute regime, the long-lived
repulsive angulon has ${B}_{1}^{\ast }=(E_{\mathrm{rep}}-E_{\mathrm{GS}})/2<B
$. When the system enters the intermediate- and high-density regimes, the
repulsive angulon state has a short lifetime and a small single-particle
residue $Z_{\mathrm{rep}}$, as indicated by the broadened repulsive angulon peak in the rotational spectrum. However, the stable attractive angulon has
a significant overlap with the rotor state, so the rotational constant ${B%
}_{1}^{\ast }=(E_{\mathrm{att}}-E_{\mathrm{GS}})/2$ should be defined with
respect to the attractive angulon energy. Since $E_{\mathrm{att}}-E_{\mathrm{%
GS}}<2B$ for all densities, it follows that $B_{1}^{\ast }<B$. The emergence of the
attractive angulon state in the intermediate density regime is crucial for
resolving the anomalous rotational constant problem. Our ansatz accounts for
the molecule's backaction on BEC, thus correctly predicting the
attractive angulon state.

\section{Conclusion}

In this work, we present a unified framework that combines an LLP-like transformation and a multireference configuration approach to address both the in- and out-of-equilibrium phenomena of a molecular impurity immersed in a superfluid bath.

We have identified two distinct types of angulon states. By accounting for the backaction of the impurity, we discovered that in the ground state, a superfluid density defect naturally forms around the molecule, leading to the emergence of a novel attractive angulon state within the defect region. This state is analogous to photonic bound states and the YSR bound state in superconductors.

Through analysis of rotational spectroscopy, we have demonstrated a crossover from the repulsive angulon to the attractive angulon as the superfluid density increases. 
In this crossover regime, the effective rotational constant satisfies ${B}^{\ast }<B$ since both the repulsive angulon in the dilute regime and the attractive angulon in all density regimes exhibit energies lower than the free rotor.

The density-dependent crossover between the repulsive and attractive angulon in rotational spectroscopy presents an avenue for validating our theoretical predictions. We predict that when broadening is observed in rotational spectroscopy, it is possible to detect a low-frequency signal associated with attractive angulons. However, in helium nanodroplets, the density is difficult to control, making it challenging to observe the crossover. While this work is motivated by molecules in superfluid environments, the underlying methods and conclusions can be broadly applicable to rotational impurities in strongly correlated, interacting environments, including molecules in ultracold atomic and molecular gases~\cite{Will2024,Jin2024,PhysRevLett.134.053001,deng2025twomanybodyphysicsultracold,zhang2025quantumphasesfinitetemperaturegases}. Ultracold atomic and molecular gases offer a high degree of tunability, enabling control over density and interaction. Dilute ultracold atomic gases provide a flexible platform for exploring physics in low- and intermediate-density regimes. Moreover, recent advances in ultracold molecular systems have enabled the realization of high-density, strongly correlated states similar to helium under specific microwave controls. We anticipate that these systems will allow for a more accessible realization of the density-dependent crossover.

\section{Acknowledgment}
We thank Richard Schmidt for the valuable discussion. Z.Z. acknowledges support from the European Union's Horizon Europe research program under the ERC Starting grant QARA (Grant No. 101041435).
This work was supported by the National Key Research and Development Program of
China (Grant No. 2021YFA0718304), by the NSFC (Grants No. 12135018 and No.
12047503, and No. 12274331).

\bibliography{library}	
\clearpage

\begin{widetext}

\appendix

\section{Equations of motion of variational parameters}
\label{EOM}

In this Appendix, we derive the EOM of variational parameters $c_{M}$ and $f_{M}(\mathbf{r})$ in the state $|\psi ^{\mathcal{J}}\rangle =\sum_{M=-\mathcal{J}}^{\mathcal{J}}c_{M}|\mathcal{J},M\rangle |f_{M}\rangle$. Here,the superposition coefficients $c_{M}$ fulfill the normalization condition $\sum_{M=-\mathcal{J}}^{\mathcal{J}}\left\vert c_{M}\right\vert ^{2}=1$, and the coherent states\begin{equation}
|f_{M}\rangle =\exp [\int d^{3}rf_{M}(\mathbf{r})\hat{\phi}^{\dagger }(\mathbf{r})-\mathrm{H.c.}]|0\rangle
\end{equation}
describes the condensate with the wavefunction $f_{M}(\mathbf{r})$.

In the imaginary- and real-time evolutions, $|\psi ^{\mathcal{J}}\rangle $
obeys
\begin{eqnarray}
\partial _{\tau }\left\vert \psi ^{\mathcal{J}}(\tau )\right\rangle &=&-%
\mathbb{P}_{\psi }[\hat{\mathcal{H}}-E(\tau )]|\psi ^{\mathcal{J}}(\tau
)\rangle ,  \label{Im} \\
i\partial _{t}\left\vert \psi ^{\mathcal{J}}(t)\right\rangle &=&\mathbb{P}%
_{\psi }\hat{\mathcal{H}}|\psi ^{\mathcal{J}}(t)\rangle ,  \label{Re}
\end{eqnarray}
where $\mathbb{P}_{\psi }$ denotes the projector onto the tangent space of the variational manifold. In the LLP frame, the Hamiltonian reads $\hat{\mathcal{H}}=\hat{\mathcal{H}}_{\mathrm{r}}+\hat{{H}}_{\mathrm{b}}+\hat{\mathcal{H}}_{\mathrm{r-b}}$, where 
\begin{eqnarray}
\hat{\mathcal{H}}_{\mathrm{r}} &=&B(\hat{\bm{\mathcal{J}}}-\hat{\mathbf{\Lambda }})^{2},  \notag \\
\hat{{H}}_{\mathrm{b}} &=&\int d^{3}r[\hat{\phi}^{\dagger }(\mathbf{r})(-\frac{\nabla ^{2}}{2m_{\mathrm{b}}}-\mu )\hat{\phi}(\mathbf{r})+\frac{g_{\mathrm{bb}}}{2}\hat{\phi}^{\dagger 2}(\mathbf{r})\hat{\phi}^{2}(\mathbf{r})],  \notag \\
\hat{\mathcal{H}}_{\mathrm{r-b}} &=&\int d^{3}r\sum_{l}V_{l}(r)Y_{l0}(\theta
,\varphi )\hat{\phi}^{\dagger }(\mathbf{r})\hat{\phi}(\mathbf{r}).
\end{eqnarray}
In the imaginary time evolution, the variational energy $E(\tau )=\langle\psi ^{\mathcal{J}}(\tau )|\hat{\mathcal{H}}|\psi ^{\mathcal{J}}(\tau)\rangle $ monotonically decreases.

It is convenient to express the operator $\hat{\phi}(\mathbf{r})=\sum_{lm}Y_{lm}(\theta ,\varphi )\hat{a}_{lm}(r)$ in the angular momentum basis, where $(r,\theta ,\varphi )$ denotes the radial coordinate, polar, and azimuthal angles of $\mathbf{r}$. The annihilation operator $\hat{a}_{lm}(r)$ in the channel $(l,m)$ satisfies the commutation relation $[\hat{a}_{lm}(r),\hat{a}_{l^{\prime }m^{\prime }}^{\dag }(r^{\prime })]=r^{-2}\delta(r-r^{\prime })\delta _{ll^{\prime }}\delta _{mm^{\prime }}$. Using the
formula
\begin{equation}
Y_{l_{1},m_{1}}(\theta ,\varphi )Y_{l_{2},m_{2}}(\theta ,\varphi )=\sum_{LM}\sqrt{\frac{(2l_{1}+1)(2l_{2}+1)}{4\pi (2l+1)}}C_{(0,0);(L,0)}^{l_{1},l_{2}}C_{(m_{1},m_{2});(L,M)}^{l_{1},l_{2}}Y_{LM}(\theta ,\varphi ),
\end{equation}
we obtain the Hamiltonian%
\begin{eqnarray}
\hat{\mathcal{H}}_{\mathrm{r}} &=&B(\hat{\bm{\mathcal{J}}}-\hat{\mathbf{%
\Lambda }})^{2},  \notag \\
\hat{{H}}_{\mathrm{b}} &=&\sum_{lm}\int r^{2}dr\hat{a}_{lm}^{\dagger }(r)(-%
\frac{\nabla _{l}^{2}}{2m_{\mathrm{b}}}-\mu )\hat{a}_{lm}(r)  \notag \\
&&+\frac{g_{\mathrm{bb}}}{2}\sum_{\{l_{i},m_{i}%
\}}G_{m_{1},m_{2},m_{3},m_{4}}^{l_{1},l_{2},l_{3},l_{4}}\int r^{2}dr\hat{a}%
_{l_{1},m_{1}}^{\dagger }(r)\hat{a}_{l_{2},m_{2}}^{\dagger }(r)\hat{a}%
_{l_{3},m_{3}}(r)\hat{a}_{l_{4},m_{4}}(r),  \notag \\
\hat{\mathcal{H}}_{\mathrm{r-b}} &=&\sum_{l,\{l_{i},m_{i}\}}\int
r^{2}drV_{l}(r)A_{0,m_{1},m_{2}}^{l,l_{1},l_{2}}\hat{a}_{l_{1},m_{1}}^{%
\dagger }(r)\hat{a}_{l_{2},m_{2}}(r)
\end{eqnarray}%
in the angular momentum basis, where $\nabla _{l}^{2}\equiv r^{-2}[\partial
_{r}(r^{2}\partial _{r})-l(l+1)]$, and%
\begin{equation}
\hat{\mathbf{\Lambda }}=\sum_{lm_{1}m_{2}}\int r^{2}dr\hat{a}%
_{l,m_{1}}^{\dagger }(r){\bm{T}}_{m_{1},m_{2}}^{[l]}\hat{a}_{l,m_{2}}(r)
\end{equation}%
is the second quantized form of the generators ${\bm{T}}%
^{[l]}=(T^{x,[l]},T^{y,[l]},T^{z,[l]})$ in the $l$-th angular momentum
channel. For instance, $T^{\alpha ,[l=0]}=0$ and 
\begin{equation*}
T^{x,[1]}=\frac{1}{\sqrt{2}}%
\begin{pmatrix}
0 & 1 & 0 \\ 
1 & 0 & 1 \\ 
0 & 1 & 0%
\end{pmatrix}%
,\ T^{y,[1]}=\frac{1}{\sqrt{2}}%
\begin{pmatrix}
0 & -i & 0 \\ 
i & 0 & -i \\ 
0 & i & 0%
\end{pmatrix}%
,\ T^{z,[1]}=%
\begin{pmatrix}
1 & 0 & 0 \\ 
0 & 0 & 0 \\ 
0 & 0 & -1%
\end{pmatrix}%
.
\end{equation*}%
The interaction coefficients%
\begin{eqnarray}
G_{m_{1},m_{2},m_{3},m_{4}}^{l_{1},l_{2},l_{3},l_{4}} &=&\sum_{L,M}\frac{%
[\prod_{i=1}^{4}(2l_{i}+1)]^{1/2}}{4\pi (2L+1)}%
C_{(0,0);(L,0)}^{l_{1},l_{2}}C_{(0,0);(L,0)}^{l_{3},l_{4}}C_{(m_{1},m_{2});(L,M)}^{l_{1},l_{2}}C_{(m_{3},m_{4});(L,M)}^{l_{3},l_{4}},
\notag \\
A_{0,m_{1},m_{2}}^{l,l_{1},l_{2}} &=&\sqrt{\frac{(2l_{2}+1)(2l+1)}{4\pi
(2l_{1}+1)}}%
C_{(0,0);(l_{1},0)}^{l_{2},l}C_{(m_{2},0);(l_{1},m_{1})}^{l_{2},l},
\end{eqnarray}%
are determined by the Clebsch-Gordan (C-G) coefficients $C_{(m_{1},m_{2});(L,M)}^{l_{1},l_{2}}=\langle l_{1},m_{1};l_{2},m_{2}|L,M,l_{1},l_{2}\rangle $. In the Edmonds convention, all C-G coefficients are real. In terms of $\hat{a}_{lm}(r)$ and $\hat{a}_{lm}^{\dag }(r)$, the coherent state is $|f_{M}\rangle =U_{M}|0\rangle $ with the displacement operator
\begin{equation}
U_{M}=\exp [\sum_{lm}\int r^{2}drf_{M,lm}(r)\hat{a}_{lm}^{\dag }(r)-\mathrm{H.c.}],
\end{equation}
where ${f}_{M,lm}(r)=\int d\Omega _{\mathbf{r}}Y_{lm}^{\ast }(\theta ,\phi ){f}_{M}(\mathbf{r})$ is the partial wave in the channel $(l,m)$.

The tangential vector in Eqs.~\eqref{Im} and \eqref{Re} is
\begin{eqnarray}
\partial _{s}|\psi ^{\mathcal{J}}\rangle &=&\sum_{M=-\mathcal{J}}^{\mathcal{J}}|\mathcal{J},M\rangle U_{M}\{[\partial _{t}c_{M}+ic_{M}\sum_{lm}\mathrm{Im}\int r^{2}drf_{M,lm}^{\ast }\left( r\right) \partial _{s}f_{M,lm}\left(r\right) ]|0\rangle  \notag \\
&&+c_{M}\sum_{lm}\int r^{2}dr\partial _{s}f_{M,lm}(r)\hat{a}_{lm}^{\dag}(r)|0\rangle \},
\end{eqnarray}
where $s=\tau $ and $t$ for the imaginary- and real-time evolutions, respectively. By projecting onto the tangent space, we obtain
\begin{eqnarray}
\partial _{\tau }c_{M} &=&-(\sum_{M^{\prime }}E_{MM^{\prime }}c_{M^{\prime}}-Ec_{M})-ic_{M}\sum_{lm}\mathrm{Im}\int r^{2}drf_{M,lm}^{\ast }\left(r\right) \partial _{\tau }f_{M,lm}\left( r\right) ,  \notag \\
c_{M}\partial _{\tau }{f}_{M,lm}\left( r\right) &=&-\sum_{M^{\prime}}\langle 0|\hat{a}_{lm}(r)U_{M}^{\dag }\hat{\mathcal{H}}_{MM^{\prime }}|{f}_{M^{\prime }}\rangle c_{M^{\prime }}
\end{eqnarray}
for the imaginary time evolution, and
\begin{eqnarray}
\partial _{t}c_{M} &=&-i\sum_{M^{\prime }}E_{MM^{\prime }}c_{M^{\prime
}}-ic_{M}\sum_{rlm}\mathrm{Im}{f}_{M,lm}^{\ast }\left( r\right) \partial _{t}%
{f}_{M,lm}\left( r\right) ,  \notag \\
c_{M}\partial _{t}{f}_{M,lm}\left( r\right) &=&-i\sum_{M^{\prime }}\langle 0|%
\hat{a}_{lm}(r)U_{M}^{\dag }\hat{\mathcal{H}}_{MM^{\prime }}|{f}_{M^{\prime
}}\rangle c_{M^{\prime }}
\end{eqnarray}%
for the real-time evolution. Here, we define $E_{MM^{\prime }}=\langle {f}_{M}|\hat{\mathcal{H}}_{MM^{\prime }}|{f}_{M^{\prime }}\rangle $ and $\hat{\mathcal{H}}_{MM^{\prime }}=\langle \mathcal{J}M|\hat{\mathcal{H}}|\mathcal{J}M^{\prime }\rangle $.

The diagonal and off-diagonal elements are
\begin{eqnarray}
E_{MM} &=&B\mathcal{J}(\mathcal{J}+1)+B\sum_{lm}l(l+1)\int r^{2}dr|{f}%
_{M,lm}(r)|^{2}+B\mathbf{\Lambda }_{M}^{2}-2B\sum_{\alpha =x,y,z}\langle 
\mathcal{J},M|\mathcal{\hat{J}}^{\alpha }|\mathcal{J},M\rangle \Lambda
_{M}^{\alpha }  \notag \\
&&+\sum_{lm}\int r^{2}dr{f}_{M,lm}^{\ast }\left( r\right) (-\frac{\nabla
_{l}^{2}}{2m_{\mathrm{b}}}-\mu ){f}_{M,lm}\left( r\right)
+\sum_{l,\{l_{i},m_{i}\}}A_{0,m_{1},m_{2}}^{l,l_{1},l_{2}}\int
r^{2}drV_{l}(r){f}_{M,l_{1}m_{1}}^{\ast }\left( r\right) {f}%
_{M,l_{2}m_{2}}\left( r\right)  \notag \\
&&+\frac{g_{\mathrm{bb}}}{2}\sum_{\{l_{i},m_{i}%
\}}G_{m_{1},m_{2},m_{3},m_{4}}^{l_{1},l_{2},l_{3},l_{4}}\int r^{2}dr{f}%
_{M,l_{1}m_{1}}^{\ast }\left( r\right) {f}_{M,l_{2}m_{2}}^{\ast }\left(
r\right) {f}_{M,l_{3}m_{3}}\left( r\right) {f}_{M,l_{4}m_{4}}\left( r\right)
,
\end{eqnarray}%
and%
\begin{equation}
E_{MM^{\prime }}=-2B\sum_{\alpha }\langle \mathcal{J},M|\mathcal{\hat{J}}%
^{\alpha }|\mathcal{J},M^{\prime }\rangle \sum_{lmm^{\prime }}\int r^{2}dr{f}%
_{M,lm}^{\ast }\left( r\right) T_{m,m^{\prime }}^{\alpha ,[l]}{f}_{M^{\prime
},lm^{\prime }}\left( r\right) \langle {f}_{M}|{f}_{M^{\prime }}\rangle ,
\end{equation}%
where ${\Lambda }_{M}^{\alpha }=\sum_{lmm^{\prime }}\int r^{2}dr{f}%
_{M,lm}^{\ast }(r)T_{mm^{\prime }}^{\alpha ,[l]}{f}_{M,lm^{\prime }}(r)$, $%
\mathbf{\Lambda }_{M}^{2}=\sum_{\alpha =x,y,z}({\Lambda }_{M}^{\alpha })^{2}$%
, and the overlap%
\begin{equation}
\langle {f}_{M}|{f}_{M^{\prime }}\rangle =\exp \{-\frac{1}{2}\sum_{lm}\int r^2dr\left[
|{f}_{M,lm}\left( r\right) |^{2}+|{f}_{M^{\prime },lm}\left( r\right) |^{2}-2%
{f}_{M,lm}^{\ast }\left( r\right) {f}_{M^{\prime },lm}\left( r\right) \right]
\}.
\end{equation}%
It follows from the commutation relation of $\hat{a}_{lm}(r)$ and $\hat{a}%
_{l,m}^{\dagger }(r)$ that

\begin{align}
\langle 0|\hat{a}_{lm}(r)U_{M}^{\dag }\hat{\mathcal{H}}_{MM}|{f}_{M}\rangle 
& = \frac{1}{r^{2}}\frac{\delta E_{MM}}{\delta {f}_{M,lm}^{\ast }\left( r\right) }, \\
\langle 0|\hat{a}_{lm}(r)U_{M}^{\dag }\hat{\mathcal{H}}_{MM^{\prime }}|{f}_{M^{\prime }}\rangle 
& = E_{MM^{\prime }}[{f}_{M^{\prime },lm}\left( r\right) -{f}_{M,lm}\left( r\right) ] \notag \\
& \quad - 2B\sum_{\alpha ,m^{\prime }}\langle \mathcal{J},M|\mathcal{\hat{J}}^{\alpha }|\mathcal{J},M^{\prime }\rangle
T_{m,m^{\prime }}^{\alpha ,[l]}{f}_{M^{\prime },lm^{\prime }}\left( r\right)
\langle {f}_{M}|{f}_{M^{\prime }}\rangle .
\end{align}

In the main text, we consider the case of the $\mathcal{J}=1$ sector, where
the generators of body rotations%
\begin{equation}
\mathcal{\hat{J}}^{x}=\frac{1}{\sqrt{2}}%
\begin{pmatrix}
0 & 1 & 0 \\ 
1 & 0 & 1 \\ 
0 & 1 & 0%
\end{pmatrix}%
,\ \mathcal{\hat{J}}^{y}=-\frac{1}{\sqrt{2}}%
\begin{pmatrix}
0 & -i & 0 \\ 
i & 0 & -i \\ 
0 & i & 0%
\end{pmatrix}%
,\ \mathcal{\hat{J}}^{z}=%
\begin{pmatrix}
1 & 0 & 0 \\ 
0 & 0 & 0 \\ 
0 & 0 & -1%
\end{pmatrix}%
.
\end{equation}%
satisfy the commutation relation $[\hat{\mathcal{J}}^{\alpha },\hat{\mathcal{J}}%
^{\beta }]=-i\epsilon _{\alpha \beta \gamma }\hat{\mathcal{J}}^{\gamma }$.

\section{Bogoliubov excitations and attractive angulon states}

\label{app: Bogo} In this Appendix, we study the Bogoliubov excitations
above the ground state, and construct the effective Hamiltonian ${H}_{%
\mathrm{eff}}$ to determine the attractive angulon state in the $\mathcal{J}%
>0$ sector.

In the $\mathcal{J}=0$ sector, the Hamiltonian in the LLP frame reads
\begin{eqnarray}
\hat{\mathcal{H}} &=&B\hat{\mathbf{\Lambda }}^{2}+\int d^{3}r\hat{\phi}%
^{\dag }(\mathbf{r})(-\frac{\nabla ^{2}}{2m_{\mathrm{b}}}-\mu )\hat{\phi}(%
\mathbf{r})  \notag \\
&&+\int d^{3}rV_{\mathrm{r-b}}(\mathbf{r})\hat{\phi}^{\dag }(\mathbf{r})\hat{%
\phi}(\mathbf{r})+\frac{g_{\mathrm{bb}}}{2}\int d^{3}r\hat{\phi}^{\dag 2}(%
\mathbf{r})\hat{\phi}(\mathbf{r})^{2}.
\end{eqnarray}
The ground state is $|\psi _{\mathrm{GS}}\rangle =|0,0\rangle |\Phi_{0}\rangle $, where
\begin{equation}
|\Phi _{0}\rangle =\exp [\int d^{3}rf_{\mathrm{GS}}(\mathbf{r})\hat{\phi}%
^{\dagger }(\mathbf{r})-\mathrm{H.c.}]|0\rangle.
\end{equation}
Using the imaginary-time evolution introduced in Appendix \ref{EOM}, we
obtain the variational parameters ${f}_{\mathrm{GS},lm}(r)=\int d\Omega _{%
\mathbf{r}}Y_{lm}^{\ast }(\Omega _{\mathbf{r}})f_{\mathrm{GS}}(\mathbf{r})$.
In the main text, we show that BEC mainly condenses in channels $(l,m)=(0,0)$ and $(1,0)$.

By introducing the fluctuation operators $\delta \hat{a}_{00}(r)=\hat{a}%
_{00}(r)-f_{\mathrm{GS},00}(r)$, $\delta \hat{a}_{10}(r)=\hat{a}_{10}(r)-f_{%
\mathrm{GS},10}(r)$, and $\hat{a}_{lm}(r)=\delta \hat{a}_{lm}(r)$ for $%
(l,m)\neq (0,0)$ and $(1,0)$, we can derive the mean-field Hamiltonian in
the $\mathcal{J}=0$ sector. For example, the mean-field Hamiltonian with the angular momentum truncation $l_{c}=1$ is as follows:
\begin{equation}
\hat{\mathcal{H}}_{\mathrm{MF}}=\frac{1}{2}\delta \alpha _{0}^{\dagger
}\left( 
\begin{array}{cc}
\mathcal{E} & \mathbf{\Delta } \\ 
\mathbf{\Delta }^{\dagger } & \mathcal{E}^{T}%
\end{array}%
\right) \delta \alpha _{0}+\delta \alpha _{1}^{\dagger }\left( 
\begin{array}{cc}
\varepsilon ^{11} & \bar{\Delta} \\ 
\bar{\Delta}^{\dagger } & \varepsilon ^{1-1}%
\end{array}%
\right) \delta \alpha _{1},
\end{equation}%
where we introduce the operators $\delta \alpha _{0}=(\delta \hat{a}_{k00},\delta \hat{a}_{k10},\delta \hat{a}_{k00}^{\dag },\delta \hat{a}_{k10}^{\dag })^{T}$ and $\delta \alpha _{1}=(\delta \hat{a}_{k11},\delta\hat{a}_{k1-1}^{\dag })^{T}$. Here, the fluctuation operators in the momentum space $\delta \hat{a}_{klm}=\sqrt{\sigma _{kl}}\int r^{2}drj_{l}(kr)\delta \hat{a}_{lm}(r)$ satisfies the commutation relation $[\delta \hat{a}_{klm},\delta \hat{a}_{klm}^{\dagger }]=1$. The spherical Bessel function $j_{l}(kr)$ satisfies
\begin{equation}
\int_{0}^{R}drr^{2}j_{l}(kr)j_{l}(k^{\prime }r)=\frac{\delta _{nn^{\prime }}%
}{\sigma _{kl}},
\end{equation}
where the discrete momentum $k=\alpha _{n}^{(l)}/R$ is determined by the zeros $\alpha _{n}^{(l)}$ of $j_{l}(x)$ and the system size $R$, and the density of the zeros
\begin{equation}
\sigma _{kl}=\frac{2}{R^{3}[j_{l+1}(\alpha _{n}^{(l)})]^{2}}.
\end{equation}
In the numerical calculation, we choose a high momentum cut-off $k_{c}=\alpha _{n_{c}}^{(l)}/R$.

The single-particle matrices
\begin{equation}
\mathcal{E}=\left( 
\begin{array}{cc}
\varepsilon ^{00} & \xi \\ 
\xi ^{\dagger } & \varepsilon ^{11}
\end{array}%
\right) ,\mathbf{\Delta }=\left( 
\begin{array}{cc}
\Delta ^{00} & \Xi \\ 
\Xi ^{T} & \Delta ^{10}
\end{array}
\right),
\end{equation}
$\varepsilon _{kk^{\prime }}^{1\pm 1}$ and $\bar{\Delta}_{kk^{\prime }}$ contain the elements
\begin{eqnarray}
\varepsilon _{kk^{\prime }}^{00} &=&(\frac{k^{2}}{2m}-\mu )\delta
_{kk^{\prime }}+\sqrt{\sigma _{k0}\sigma _{k^{\prime }0}}\int
r^{2}drj_{0}(kr)[A_{1}V_{0}(r)+2g_{bb}(G_{1}f_{\mathrm{GS}%
,00}^{2}(r)+G_{2}f_{\mathrm{GS},10}^{2}(r))]j_{0}(k^{\prime }r),  \notag \\
\varepsilon _{kk^{\prime }}^{10} &=&[2B+(\frac{k^{2}}{2m}-\mu )]\delta
_{kk^{\prime }}+\sqrt{\sigma _{k1}\sigma _{k^{\prime }1}}\int
r^{2}drj_{1}(kr)[A_{5}V_{0}(r)+2g_{bb}(G_{2}f_{\mathrm{GS}%
,00}^{2}(r)+G_{17}f_{\mathrm{GS},10}^{2}(r))]j_{1}(k^{\prime }r)  \notag \\
\varepsilon _{kk^{\prime }}^{1\pm 1} &=&[2B+(\frac{k^{2}}{2m}-\mu )]\delta
_{kk^{\prime }}+2B\sqrt{\sigma _{k1}\sigma _{k^{\prime }1}}\int
r^{2}drj_{1}(kr)f_{\mathrm{GS},10}(r)\int r^{\prime 2}dr^{\prime
}j_{1}(k^{\prime }r^{\prime })f_{\mathrm{GS},10}(r^{\prime })  \notag \\
&&+\sqrt{\sigma _{k1}\sigma _{k^{\prime }1}}\int
r^{2}drj_{1}(kr)[A_{4}V_{0}(r)+2g_{bb}(G_{3}f_{\mathrm{GS}%
,00}^{2}(r)+G_{21}f_{\mathrm{GS},10}^{2}(r))]j_{1}(k^{\prime }r)  \notag \\
\xi _{kk^{\prime }} &=&\sqrt{\sigma _{k0}\sigma _{k^{\prime }1}}\int
r^{2}drj_{0}(kr)[A_{2}V_{1}(r)+2g_{bb}(G_{2}+G_{5})f_{\mathrm{GS},00}(r)f_{
\mathrm{GS},10}(r)]j_{1}(k^{\prime }r),
\end{eqnarray}
and
\begin{eqnarray}
\Delta _{kk^{\prime }}^{00} &=&g_{bb}\sqrt{\sigma _{k0}\sigma _{k^{\prime }0}%
}\int r^{2}drj_{0}(kr)(G_{1}f_{\mathrm{GS},00}^{2}(r)+G_{5}f_{\mathrm{GS}%
,10}^{2}(r))j_{0}(k^{\prime }r),  \notag \\
\Delta _{kk^{\prime }}^{10} &=&g_{bb}\sqrt{\sigma _{k1}\sigma _{k^{\prime }1}%
}\int r^{2}drj_{1}(kr)(G_{5}f_{\mathrm{GS},00}^{2}(r)+G_{17}f_{\mathrm{GS}%
,10}^{2}(r))j_{1}(k^{\prime }r),  \notag \\
\Xi _{kk^{\prime }} &=&2g_{bb}\sqrt{\sigma _{k0}\sigma _{k^{\prime }1}}\int
r^{2}drj_{0}(kr)G_{2}f_{\mathrm{GS},00}(r)f_{\mathrm{GS},10}^{2}(r)j_{1}(k^{%
\prime }r),  \notag \\
\bar{\Delta}_{kk^{\prime }} &=&2B\sqrt{\sigma _{k1}\sigma _{k^{\prime }1}}%
\int r^{2}drj_{1}(kr)f_{\mathrm{GS},10}(r)\int r^{\prime 2}dr^{\prime
}j_{1}(k^{\prime }r^{\prime })f_{\mathrm{GS},10}(r^{\prime })  \notag \\
&&+g_{bb}\sqrt{\sigma _{k1}\sigma _{k^{\prime }1}}\int
r^{2}drj_{1}(kr)(G_{6}f_{\mathrm{GS},00}^{2}(r)+G_{18}f_{\mathrm{GS}%
,10}^{2}(r))j_{1}(k^{\prime }r).
\end{eqnarray}%
Here, we abbreviate the notations as
\begin{equation}
A_{0,0,0}^{0,0,0}=A_{1},A_{0,0,0}^{1,0,1}=A_{2},A_{0,0,0}^{1,1,0}=A_{3},A_{0,1,1}^{0,1,1}=A_{4},A_{0,0,0}^{0,1,1}=A_{5},
\notag \\
A_{0,-1,-1}^{0,1,1}=A_{6},A_{0,1,1}^{1,1,1}=A_{7},A_{0,0,0}^{1,1,1}=A_{8},A_{0,-1,-1}^{1,1,1}=A_{9},
\end{equation}
and
\begin{equation}
G_{0,0,0,0}^{0,0,0,0}=G_{1},G_{0,0,0,0}^{1,0,1,0}=G_{2},G_{1,0,1,0}^{1,0,1,0}=G_{3},G_{-1,0,-1,0}^{1,0,1,0}=G_{4},G_{0,0,0,0}^{1,1,0,0}=G_{5},G_{1,-1,0,0}^{1,1,0,0}=G_{6},\notag
\end{equation}%
\begin{equation}
G_{0,0,0,0}^{0,0,1,1}=G_{7},G_{0,0,1,-1}^{0,0,1,1}=G_{8},G_{0,0,0,0}^{1,1,1,0}=G_{9},G_{1,-1,0,0}^{1,1,1,0}=G_{10},G_{1,0,1,0}^{1,1,1,0}=G_{11},G_{-1,0,-1,0}^{1,1,1,0}=G_{12},
\notag
\end{equation}
\begin{equation}
G_{0,0,0,0}^{1,0,1,1}=G_{13},G_{0,0,1,-1}^{1,0,1,1}=G_{14},G_{1,0,1,0}^{1,0,1,1}=G_{15},G_{-1,0,-1,0}^{1,0,1,1}=G_{16},G_{0,0,0,0}^{1,1,1,1}=G_{17},G_{1,-1,0,0}^{1,1,1,1}=G_{18},
\notag
\end{equation}
\begin{equation}
G_{0,0,1,-1}^{1,1,1,1}=G_{19},G_{1,-1,1,-1}^{1,1,1,1}=G_{20},G_{1,0,1,0}^{1,1,1,1}=G_{21},G_{-1,0,-1,0}^{1,1,1,1}=G_{22},G_{1,1,1,1}^{1,1,1,1}=G_{23},G_{-1,-1,-1,-1}^{1,1,1,1}=G_{24},
\end{equation}
where $A_{7\sim 9}$ and $G_{9\sim 16}$ are zero.

The Hamiltonian can be diagonalized via $\delta \alpha_{m=0,1}=S_{m=0,1}\beta _{m=0,1}$, where $\beta _{0}=(\hat{b}_{\alpha 0}, \hat{b}_{\alpha 0}^{\dag })^{T}$ is determined by $2n_{c}$-dimensional vector $\hat{b}_{\alpha =1,...2n_{c},0}$ and $\beta _{1}=(\hat{b}_{\alpha 1}, \hat{b}_{\alpha -1}^{\dag })^{T}$ is determined by $n_{c}$-dimensional vectors $\hat{b}_{\alpha =1,...,n_{c},\pm 1}$. The Bogoliubov transformations are
\begin{equation}
S_{m}=
\begin{pmatrix}
\mathcal{U}^{(m)} & \mathcal{V}^{(m)\ast } \\ 
\mathcal{V}^{(m)} & \mathcal{U}^{(m)\ast }
\end{pmatrix},
\end{equation}
where
\begin{equation*}
\mathcal{U}^{(0)}=
\begin{pmatrix}
\mathcal{U}_{11}^{0} & \mathcal{U}_{12}^{0} \\ 
\mathcal{U}_{21}^{0} & \mathcal{U}_{22}^{0}
\end{pmatrix}
,\mathcal{V}^{(0)}=
\begin{pmatrix}
\mathcal{V}_{11}^{0} & \mathcal{V}_{12}^{0} \\ 
\mathcal{V}_{21}^{0} & \mathcal{V}_{22}^{0}%
\end{pmatrix}
\end{equation*}
are $2n_{c}\times 2n_{c}$ matrices formed by $n_{c}\times n_{c}$ blocks $\mathcal{U}_{ij}^{0}$ and $\mathcal{V}_{ij}^{0}$, and $\mathcal{U}^{(1)}$ and $\mathcal{V}^{(1)}$ are $n_{c}\times n_{c}$ matrices. The transformations satisfy
\begin{equation}
S_{m}^{\dagger }%
\begin{pmatrix}
1 & 0 \\ 
0 & -1%
\end{pmatrix}%
S_{m}=%
\begin{pmatrix}
1 & 0 \\ 
0 & -1%
\end{pmatrix}%
,
\end{equation}%
and diagonalize the single-particle matrices as%
\begin{eqnarray}
S_{0}^{\dagger }\left( 
\begin{array}{cc}
\mathcal{E} & \mathbf{\Delta } \\ 
\mathbf{\Delta }^{\dagger } & \mathcal{E}^{T}%
\end{array}%
\right) S_{0} &=&\left( 
\begin{array}{cc}
E_{0} & 0 \\ 
0 & E_{0}%
\end{array}%
\right) ,  \notag \\
S_{1}^{\dagger }\left( 
\begin{array}{cc}
\varepsilon ^{11} & \bar{\Delta} \\ 
\bar{\Delta}^{\dagger } & \varepsilon ^{1-1}%
\end{array}%
\right) S_{1} &=&\left( 
\begin{array}{cc}
E_{1} & 0 \\ 
0 & E_{1}%
\end{array}%
\right),
\end{eqnarray}%
where the diagonal matrix $E_{0}$ has the element $E_{\alpha
=1,...,2n_{c},0} $, and the diagonal matrix $E_{1}$ has the element $%
E_{\alpha =1,...,n_{c},1} $. In terms of the Bogoliubov operators $\beta
_{m} $, the mean-field Hamiltonian reads%
\begin{equation}
\hat{\mathcal{H}}_{\mathrm{MF}}=\sum_{\alpha =1}^{2n_{c}}E_{\alpha 0}\hat{b}%
_{\alpha 0}^{\dag }\hat{b}_{\alpha 0}+\sum_{\alpha =1}^{n_{c}}E_{\alpha
1}\sum_{m=\pm 1}\hat{b}_{\alpha m}^{\dag }\hat{b}_{\alpha m}.
\end{equation}

In the $\mathcal{J}=1$ sector, we project the Hamiltonian in the subspace $%
\mathcal{S}=\{|\Xi ^{\mathcal{J}=1}\rangle ,\left\vert \mathcal{J}%
,m\right\rangle \hat{b}_{\alpha m}^{\dag }|\Phi _{0}\rangle \}$. For the
truncation $l_{c}=1$, the effective Hamiltonian reads%
\begin{equation}
H_{\mathrm{eff}}=%
\begin{pmatrix}
2B & 0 & -2B\lambda ^{\lbrack 1]\ast } & -2B\lambda ^{\lbrack 1]} \\ 
0 & E_{0}+2B & -2B\lambda ^{{[2]}\dag } & -2B\lambda ^{\lbrack 3]\dagger }
\\ 
-2B\lambda ^{\lbrack 1]} & -2B\lambda ^{\lbrack 2]} & E_{1} & 0 \\ 
-2B\lambda ^{\lbrack 1]\ast } & -2B\lambda ^{\lbrack 3]} & 0 & E_{1}%
\end{pmatrix}%
.
\end{equation}%
in the basis $\{|\Xi ^{1}\rangle ,\left\vert 1,0\right\rangle \hat{b}%
_{\alpha 0}^{\dag }|\Phi _{0}\rangle ,\left\vert 1,1\right\rangle \hat{b}%
_{\alpha 1}^{\dag }|\Phi _{0}\rangle ,\left\vert 1,-1\right\rangle \hat{b}%
_{\alpha -1}^{\dag }|\Phi _{0}\rangle \}$, where%
\begin{eqnarray}
\lambda _{\alpha }^{[1]} &=&\sum_{k}(\mathcal{U}^{1\dag }+\mathcal{V}%
^{1T})_{\alpha k}[\sqrt{\sigma _{k1}}\int r^{2}drj_{1}(kr){f}_{\mathrm{GS}%
,10}(r)],  \notag \\
\lambda ^{\lbrack 2]} &=&\left( 
\begin{array}{cc}
\mathcal{U}^{1\dag } & \mathcal{V}^{1T}%
\end{array}%
\right) \left( 
\begin{array}{cc}
\mathcal{U}_{21}^{0} & \mathcal{U}_{22}^{0} \\ 
\mathcal{V}_{21}^{0} & \mathcal{V}_{22}^{0}%
\end{array}%
\right) ,\lambda ^{\lbrack 3]}=(\mathcal{U}^{1T},\mathcal{V}^{1\dagger
})\left( 
\begin{array}{cc}
\mathcal{U}_{21}^{0} & \mathcal{U}_{22}^{0} \\ 
\mathcal{V}_{21}^{0} & \mathcal{V}_{22}^{0}%
\end{array}%
\right).
\end{eqnarray}

By diagonalizing $H_{\mathrm{eff}}$ numerically, we obtain the excited states and their spectrum in the $\mathcal{J}=1$ sector. The eigenstate
\begin{equation}
|\psi _{n}^{\mathcal{J}}\rangle =\sqrt{Z_{n}}|\Xi ^{\mathcal{J}}\rangle
+\sum_{\alpha m}\psi _{n,\alpha m}\left\vert \mathcal{J},m\right\rangle \hat{%
b}_{\alpha m}^{\dag }|\Phi _{0}\rangle
\end{equation}%
takes the same form as the Chevy ansatz, except that the background $|\Phi _{0}\rangle $ is non-uniform. The non-uniform condensate plays a critical role by creating a trapping potential for excitations with finite angular momentum, capable of supporting a bound state when the trap is sufficiently deep. Furthermore, it provides a better alignment with the rotational spectroscopy experiment.

\end{widetext}

\end{document}